\documentclass[aps,prd,twocolumn,groupedaddress]{revtex4}

\usepackage{enumitem}
\usepackage{graphicx}
\usepackage{dcolumn}
\usepackage{array,multirow}
\usepackage{bm}
\usepackage{amsmath}
\usepackage{epstopdf}
\usepackage{amsfonts}
\usepackage{amssymb}
\usepackage{epsfig}
\usepackage{tabularx}
\usepackage{color}
\usepackage{xcolor}
\usepackage[colorlinks=true,citecolor=blue,urlcolor=cyan,breaklinks]{hyperref}

\newcommand{\bea}{\begin{eqnarray}}
\newcommand{\ena}{\end{eqnarray}}

\def\be{\begin{equation}} 
\def\ee{\end{equation}}   

\begin{document}

\title{Scale-dependent rotating BTZ black hole}

\author{\'Angel Rinc\'on}
\email{arrincon@uc.cl}
\affiliation{Instituto de F\'{i}sica, Pontificia Universidad Cat\'{o}lica de Chile, \\ \mbox{Avenida Vicu\~na Mackenna 4860, Santiago, Chile.}}

\author{Benjamin Koch}
\email{bkoch@fis.puc.cl}
\affiliation{Instituto de F\'{i}sica, Pontificia Universidad Cat\'{o}lica de Chile, \\ \mbox{Avenida Vicu\~na Mackenna 4860, Santiago, Chile.}}

\date{\today}

\begin{abstract}
This work presents a generalization of the rotating black hole in two plus one dimensions, in the light of scale--dependent gravitational couplings. In particular, the gravitational coupling $\kappa_0$ and the cosmological term $\Lambda_0$ are not forced to be constants anymore. Instead, $\kappa$ and $\Lambda$ are allowed to change along the radial scale $r$. The effective Einstein field equations of this problem are solved by assuming static rotational symmetry and by maintaining the usual structure of the line element. For this generalized solution, the asymptotic behavior, the horizon structure, and the thermodynamic properties are analyzed.
\end{abstract}


\maketitle


\section{Introduction}

To formulate a consistent and predictive quantum theory of gravity (QG) is one of the mayor challenges for the community seeking a unified description of the known fundamental interactions. Currently, at least 16 major approaches to quantum gravity have been proposed in the literature (see \cite{Esposito:2011rx} and references therein), but none of these approaches have reached the goal in a completely satisfactory way. 

In this paper we contribute to the topic of quantum gravity by
studying black hole solutions
of effective scale--dependent gravity in $2+1$ dimensions.
We thus, combine three different aspects, namely, scale dependence,
gravity in $2+1$ dimensions and black holes. Each of those aspects hast a motivation of its own, 
but all of those aspects have an important motivation from the perspective of quantum gravity:

\begin{itemize}
\item[$\bullet$] Black holes ({\bf BH}s):\\
Black Holes are objects of paramount importance in gravitational theories \cite{Chandrasekhar:1985kt}. 
They allow to study gravitational systems at the
transition between a quantum and a classical regime as for example through the 
 the famously predicted Hawking radiation \cite{Hawking:1974rv,Hawking:1974sw}.
BHs are thus excellent laboratories to investigate and understand several aspects of 
general relativity
at the transition between a classical and quantum regime \cite{Calmet:2015fua}.

\item[$\bullet$]  $2+1$ dimensions:\\
It can be expected that the features of a successful solution of the problem of quantum gravity are universal for gravitational theories of different dimensionality. 
Since gravity in $2+1$ dimensions is mathematically less involved than in $3+1$ dimensions, this 
lower dimensional theory is a good toy model if one aims to understand the underlying mechanisms of
full quantum gravity in $3+1$ dimensions.
Apart from this motivation by quantum gravity,
the study of gravity in $2+1$ dimensions is of interest because of its 
deep connection to Chern-Simons theory \cite{Witten:1988hc,Witten:2007kt} and because of its applications in the context of the AdS/CFT correspondence \cite{Maldacena:1997re,Strominger:1997eq,
Balasubramanian:1999re,Aharony:1999ti}.
Within this lower dimensional gravitation theory 
the black hole solution found by Ba\~nados, Teitelboim, and Zanelli (BTZ) \cite{Banados:1992wn,Banados:1992gq} 
plays a crucial role.

\item[$\bullet$] Scale dependence ({\bf{SD}}):\\
Before actually attacking the whole problem of QG with all its different, and up to now limited, realizations, 
one can begin with a more modest approach and concentrate on generic common features, which are expected
from such a theory.
One feature which is shared by most of the candidate theories for quantum gravity (actually by most quantum field theories) 
is that they predict a scale dependence of the coupling constants in the corresponding effective action.
Luckily there is a well defined formalism which allows to deduce background solutions
from a given effective action.
We will follow those techniques which have been previously probed 
with a variety of problems
\cite{Koch:2010nn,Contreras:2013hua,
Koch:2014joa,
Koch:2015nva,
Contreras:2016mdt,
Koch:2016uso,
Rincon:2017ypd,
Rincon:2017goj,
Rincon:2017ayr,
Contreras:2017eza,
Hernandez-Arboleda:2018qdo,
Contreras:2018dhs,
Rincon:2018sgd,
Contreras:2018swc}.
In this paper we aim to study the dominant effects such a scale dependence could have on the BTZ black hole in the Einstein Hilbert truncation of the  effective action of gravity in 2+1 dimensions.
By using a well defined method which is based on the variational principle one can 
explore leading local effects of quantum gravity on a rotationally symmetric space-time in a source free region (like BTZ), 
even without the knowledge of the exact underlying theory.
\end{itemize}

\begin{figure}[ht]
  \begin{picture}(370,235)
  \includegraphics[width=\linewidth]{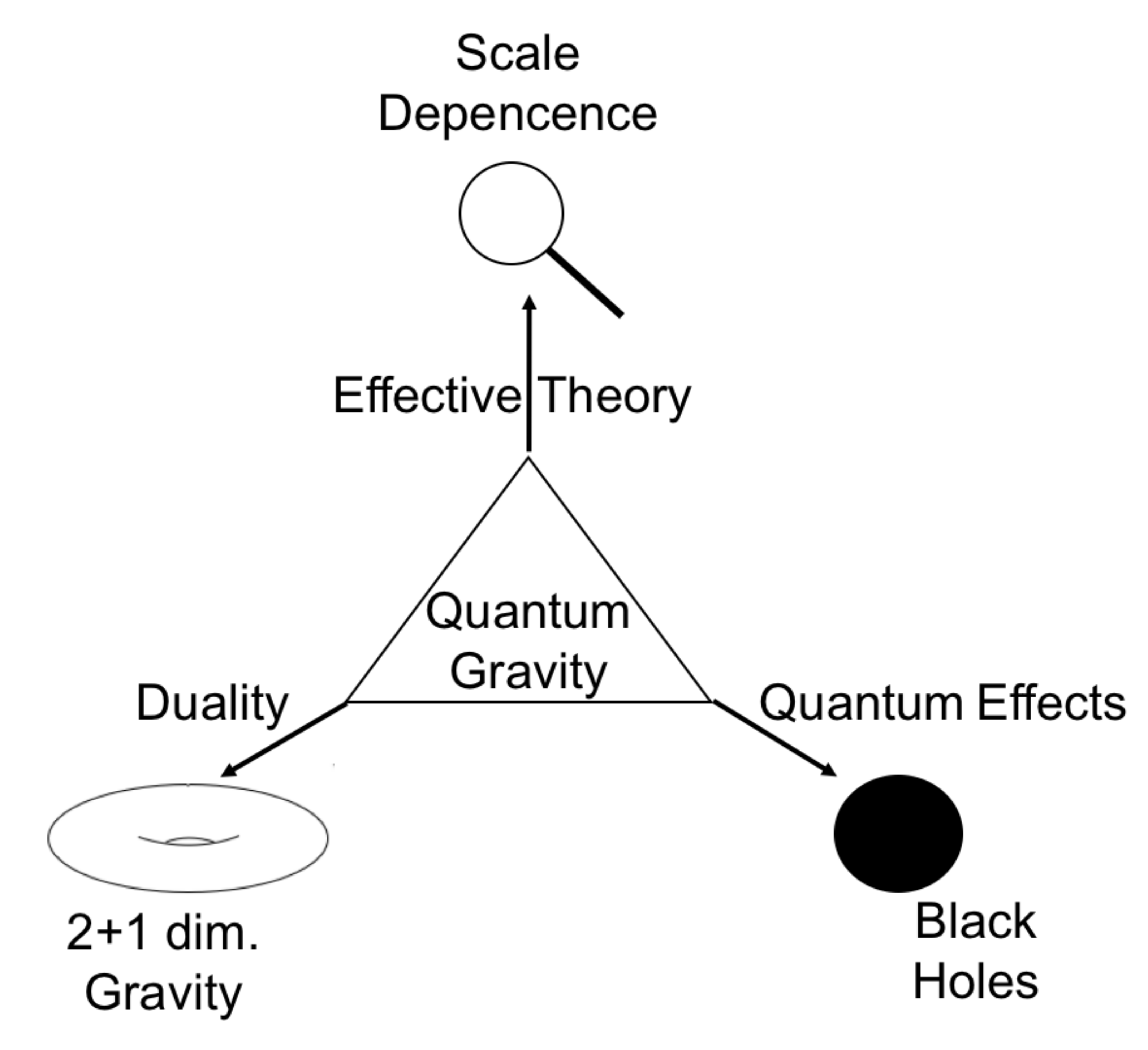}
  \end{picture}    
 \caption{\label{figscheme} 
Conceptual flow chart for the interplay of SD, BHs, and $2+1$ dimensions with QG.}
\end{figure}

The important connection of those three ingredients with the underlying topic
of QG is shown in figure~\ref{figscheme},
showing clearly that the study of corrections to the classical BTZ solution,
as those derived in this paper, are a key test for any theory of QG.

This paper is organized as follows: 
after this introduction, we present the action and the classical BTZ solution in the next section.
Then, the general framework of this work is introduced in section \ref{SDCSS}. 
The scale dependence for a rotating BTZ black hole is presented in section  \ref{SDRBTZ}. The bevaviour of the Ricci scalar, the asymptotic space-time as well as the thermodynamics is investigated in Sect. \ref{Invariants} and \ref{Thermodynamic} respectively. The discussion of this result and remarks are shown in section \ref{Discussion}.
 The main ideas and results are summarized in the conclusion section  \ref{Conclusion}.
 Note that throughout the paper we will use
natural units with ($c= \hbar = k_B = 1$).

\section{Classical BTZ solution with $J_0 \neq 0$}

This section reminds of some key features of the classical BTZ black hole solution \cite{Banados:1992wn,Banados:1992gq},
such as line element, event horizons, and thermodynamics.
Besides, the contribution of angular momentum will be consi\-dered focussing on the extremal black hole case. The mi\-nimal coupling between gravity and matter is described by the the Einstein Hilbert action
\be\label{actionBTZ}
I_0[g_{\mu \nu}] = \int {\mathrm {d}}^3x \sqrt{-g} \Bigg[\frac{1}{2 \kappa_0}\Bigl(R - 2\Lambda_0\Bigl)  + \mathcal{L}_{M} \Bigg],
\ee
where $g_{\mu \nu}$ is the metric field, $R$ is the Ricci scalar, $\kappa_0 \equiv 8 \pi G_0$ is the gravitational coupling, $\Lambda_0$ is the cosmological constant, $\mathcal{L}_{M}$ is the matter Lagrangian, and $g$ is the determinant of the metric field. 
The classical Einstein field equations are obtained from (\ref{actionBTZ}) by varying the action with respect to the metric field
\be\label{eomBTZ}
 G_{\mu\nu} + \Lambda_{0} g_{\mu\nu} = \kappa_0 T_{\mu \nu},
\ee
where $T_{\mu \nu}$ is the energy momentum tensor associated to a matter source
\begin{align}
T_{\mu \nu} &\equiv T^{M}_{\mu \nu} = -2 \frac{\delta \mathcal{L}_{M}}{\delta g^{\mu \nu}} + \mathcal{L}_M g_{\mu \nu} .
\end{align}
For the case of rotational symmetry without any matter contribution, the metric 
solution of (\ref{eomBTZ})
takes the form
\be\label{lineele}
\mathrm{d}s^2= -f_0(r) \mathrm{d}t^2+ f_0(r)^{-1} \mathrm{d}r^2 + r^2 \Bigl[N_0(r)\mathrm{d}t + \mathrm{d}\phi \Bigl]^2.
\ee
Here, $f_0(r)$ and $N_0(r)$ are the lapse function and the shift function respectively,
which are given by
\begin{align}\label{classical}
f_0(r) &= - 8 M_0 G_0 + \frac{r^2}{\ell_0^2} + \frac{16 G_0^2 J_0^2}{r^2},
\\
N_0(r) &= - \frac{4 G_0 J_0}{r^2},
\end{align}
where $\ell_0$ is defined by $\Lambda_0 \equiv -1/\ell_0^2$.
The two constants of integration $M_0$ and $J_0$ 
are the conserved charges associated to asymptotic invariance under time shifts (mass) and rotations (angular momentum) respectively.
The horizons
\begin{align}\label{horizon_Classic}
(r_0^{\pm})^2 &= 4 G_0 M_0 \ell_0^2 \Bigl[1 \pm \Delta \Bigl],
\end{align}
 are defined through the condition $f(r_0^\pm)=0$. 
Here, the parameter $\Delta$ encodes the impact of the rotational contribution on the event horizon
\begin{align}\label{Delta}
\Delta &= \sqrt{1 -\Bigg( \frac{J_0}{M_0 \ell_0}\Bigg)^2}.
\end{align}
The positive root $r_0^+$ is the black hole's outer horizon. 
One can express the lapse function in terms of the event horizons
\begin{align}
f_0(r) &= \frac{1}{\ell_0^2 r^2}\Bigg[ \Bigl(r^2 - (r_{0}^{+})^2\Bigl) \Bigl(r^2 - (r_{0}^{-})^2\Bigl)\Bigg].
\end{align}
It is important to note that, the parameters must satisfy
\begin{align}
M_0 > 0, \hspace{0.5cm} \wedge \hspace{0.5cm} |J_0| \leq M_0 \ell_0,
\end{align}
in order to get physical solutions. 
When the classical angular momentum takes a maximum value given by
\begin{align}\label{Jcritico}
J_{0}^{\ \text{max}} = M_0 \ell_0,
\end{align}
the solution is called an extremal black hole.
Regarding black hole thermodynamics, the Bekenstein-Hawking entropy is given by
\begin{align}
S_0(r_0^+) &= \frac{\mathcal{A}_H(r_0^{+})}{4G_0}.
\end{align}
The corresponding Hawking temperature is
\begin{align} \label{Tclasica}
T_0(r_0^+) &= \frac{1}{4\pi} 
\Bigg|
\frac{16 G_0 M_0 }{r_0^+} \Delta
\Bigg|,
\end{align}
where $\mathcal{A}_H(r_0)$ is the horizon area which is given by
\begin{align}
\mathcal{A}_{H}(r_0^+) &= \oint  {\mathrm {d}}x \sqrt{h} = 2 \pi r_0^+.
\end{align}

\section{Scale dependent couplings and scale setting} \label{SDCSS}

This section resumes the  implementation of scale dependence that was used for the present work. 
The notation and procedures follow 
\cite{Koch:2010nn,Contreras:2013hua,
Koch:2014joa,
Koch:2015nva,
Contreras:2016mdt,
Koch:2016uso,
Rincon:2017ypd,
Rincon:2017goj,
Rincon:2017ayr,
Contreras:2017eza,
Hernandez-Arboleda:2018qdo,
Contreras:2018dhs,
Rincon:2018sgd,
Contreras:2018swc,
Reuter:2003ca,
Domazet:2012tw}.
In this framework the scale dependence is implemented at the level of an effective action
as a generalization of the classical action.
For the case of (\ref{actionBTZ}), the truncated effective action takes the form 
\be\label{actionEG}
\Gamma[g_{\mu \nu}, k] = \int {\mathrm {d}}^3x \sqrt{-g}
\Bigg[\frac{1}{2\kappa_{k}}\Bigl(R-2\Lambda_k\Bigl)
+ \mathcal{L}_M
\Bigg].
\ee
As shown in \cite{Koch:2016uso}, this action is consistent at the classical 
level if one sets the arbitrary scale based on a variational principle, which means
that the scale $k$ considered as a non-dynamical field instead of a global constant. 
A variation of (\ref{actionEG}) with respect to the metric field $g_{\mu \nu}$ gives the modified Einstein equations
\begin{eqnarray}
 G_{\mu\nu} + g_{\mu\nu}\Lambda_{k} = \kappa_k T^{\text{effec}}_{\mu \nu}.
\label{eomg}
\end{eqnarray}
Here, the effective stress energy tensor is defined as
\begin{align}
\kappa_k T^{\text{effec}}_{\mu\nu} &= \kappa_k T^{M}_{\mu\nu} -  \Delta t_{\mu \nu},
\end{align}
which consists of the usual stress energy of the matter Lagrangian $T^{M}_{\mu\nu}$
and an additional contribution due to the scale dependence of the gravitational coupling
\begin{align}\label{deltaT}
\Delta t_{\mu\nu} &= G_k \Bigl(g_{\mu \nu} \square - \nabla_{\mu} \nabla_{\nu}
\Bigl)G_k^{-1}.
\end{align}
For the vacuum solution presented in this paper, the pure matter contribution is absent $T^{M}_{\mu\nu}=0$.

Varying the action (\ref{actionEG}) with respect to the scale-field $k(x)$ gives
\be\label{eomk}
\left[R \frac{\partial}{\partial k} \left(\frac{1}{G_k}\right)-
2 \frac{\partial}{\partial k}\left(\frac{\Lambda_k}{G_k}\right)\right]\cdot \partial k
=0.
\ee
The above equations of motion are consistently complemented
by the Bianchi identity, reflecting invariance under 
coordinate transformations
\be\label{diffeo}
\nabla^\mu G_{\mu \nu}=0.
\ee

\section{Scale dependence BTZ solution with $J_0 \neq 0$} \label{SDRBTZ}

The line element consistent with a static space-time, with rotational symmetry
is given by
\be\label{lineelegen}
\mathrm{d}s^2= -f(r) \mathrm{d}t^2 + g(r) \mathrm{d}r^2 + r^2 \Bigl[N(r)\mathrm{d}t + \mathrm{d}\phi\Bigl]^2,
\ee
where $f(r)$, $g(r)$ $N(r)$, and $k(r)$ are functions that must be determined from the equations
of motion ~(\ref{eomg}-\ref{diffeo}).
When the functional scale dependence of the couplings $G_k$ and $\Lambda_k$
is known, the system closes into itself and the equations (\ref{eomg}-\ref{diffeo}) allow,
at least numerically to determine the functions $f(r)$, $g(r)$ $N(r)$, and $k(r)$ \cite{Koch:2014joa}.
In certain truncations and functional approaches such as the
functional renormalization group approach it is indeed possible
to study scale dependence and approximate improvement
of classical black hole solutions \cite{Bonanno:1998ye,
Bonanno:2000ep,
Bonanno:2006eu,
Reuter:2006rg,
Falls:2012nd,
Cai:2010zh,
Becker:2012js,
Becker:2012jx,
Koch:2013owa,
Koch:2013rwa,
Ward:2006vw,
Burschil:2009va,
Falls:2010he,
Koch:2014cqa,
Bonanno:2016dyv}.
However, those approximation scenarios are subject to theoretical uncertainties
related to the truncations used to calculate the beta functions. Further, due to the implicit assumption of 
improvement of classical solutions, they typically
do not solve the whole selfconsistent system of equations ~(\ref{eomg}-\ref{diffeo}) anymore.

The idea is to avoid 
the theoretical uncertainties inflicted with the usage of given functions  $G_k$ and $\Lambda_k$, and instead
to learn about the radial dependence of the functions $G(r)$ and $\Lambda(r)$
directly from the selfconsistent system of equations ~(\ref{eomg}-\ref{diffeo}).
Thus, instead of trying to solve for the four functions $\left\{f(r), \, g(r),\, N(r),\, k(r)\right\}$ for given, but uncertain,  $G_k$ and $\Lambda_k$
one can try to solve the equations (\ref{eomg}-\ref{diffeo}) directly for the five functions $\left\{f(r),\, g(r),\, \Lambda(r),\, G(r),\, N(r)\right\}$.
Here, $G(r)$ and $\Lambda(r)$ have inherited their radial dependence from $k(r)$.
The problem for this elegant workaround is that there are now five unknown functions in a
system which only has four independent equations.
Thus, one additional condition is needed in order to be able to fully solve this system of equations.
Following previous findings \cite{Koch:2010nn,Domazet:2012tw,Koch:2014joa,
Contreras:2016mdt,Rincon:2017goj,Rincon:2017ypd} this additional condition is
that we restrict to solutions which fulfill the so-called Schwarzschild relation, namely that $g(r) \equiv f(r)^{-1}$.
Therefore, the corresponding line element is 
\begin{align}\label{lineelans}
\mathrm{d}s^2 &= -f(r) \mathrm{d}t^2 + f(r)^{-1} \mathrm{d}r^2 + r^2 \Bigl[N(r)\mathrm{d}t + \mathrm{d}\phi\Bigl]^2
\end{align}
and the equations of motion can be solved for the four functions
$\left\{f(r), \, \Lambda(r),\, G(r),\, N(r)\right\}$.

\subsection{Solution}
\label{sec_findBTZ}

\begin{figure}[ht]
  \begin{picture}(370,235)
 
  \includegraphics[width=\linewidth]{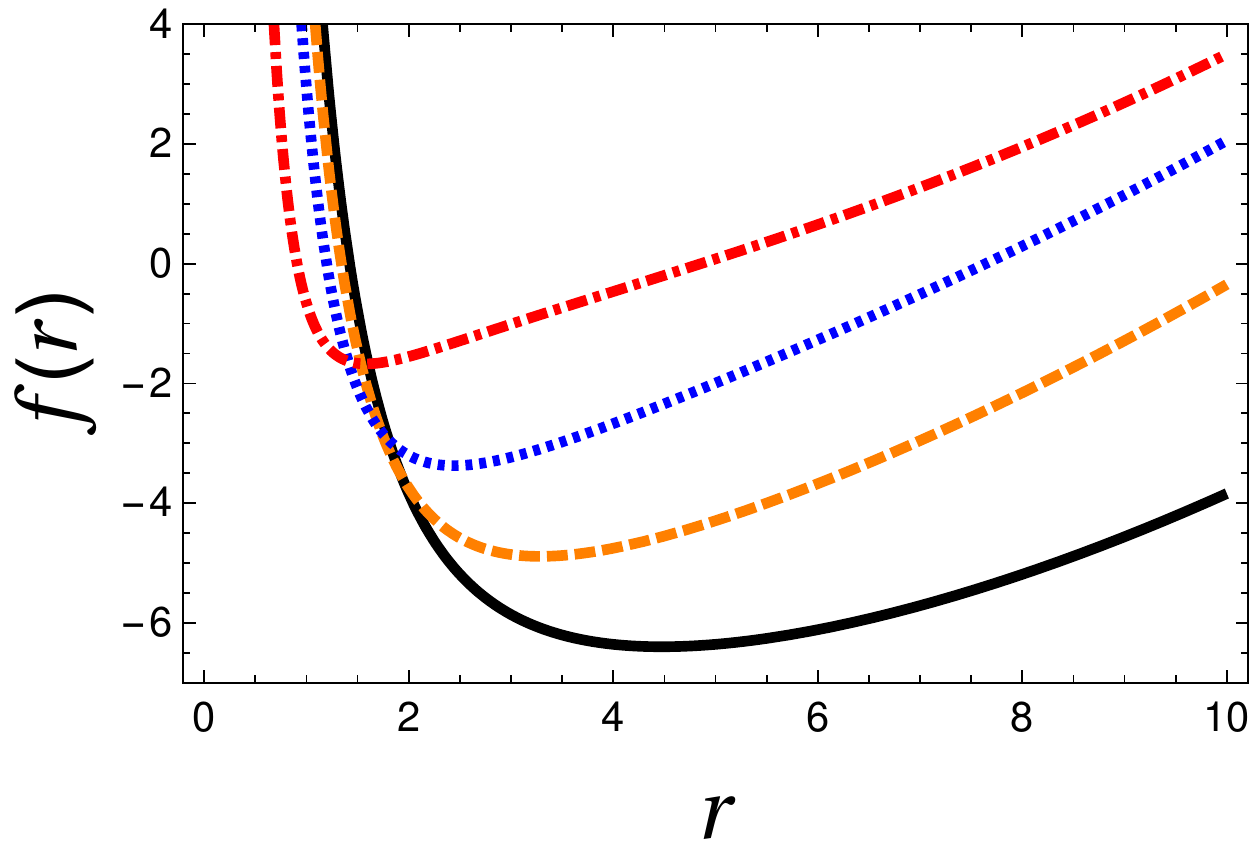}

  \end{picture}    
      \caption{\label{figfr} 
    Radial dependence of the lapse function $f(r)$ for $\ell_0=5$, $G_0=1$, $M_0=1${\textcolor{blue}{,}}
     and $J_0 = 1$.
The different curves correspond to 
the classical case 
$\epsilon=0$ 
solid black line, 
$\epsilon=0.05$ 
dashed orange line,
$\epsilon=0.2$ 
dotted blue line{\textcolor{blue}{,}}
and
$\epsilon=1$ 
dot-dashed red line.
}
\end{figure}

Based on the ansatz~(\ref{lineelans}) one finds that the equations~(\ref{eomg}) are solved by
\begin{align}\label{setsoluII}
G(r) =& \frac{G_0}{1 + r \epsilon},
\\
N(r) =& - \frac{4 G_0 J_0}{r^2} Y(r),
\\ \label{solf}
f(r) =& -8 M_0 G_0 Y(r) + \frac{r^2}{\ell_0^2} + \frac{16 G_0^2 J_0^2}{r^2}Y(r)^2,
\\
\Lambda(r) =& - \frac{r + 3r^2 \epsilon - 8 G_0 \ell_0^2 M_0 \epsilon Y(r) }{\ell_0^2 r (1 + r \epsilon)}
-
\frac{4 G_0^2 J_0^2}{r^2}Y(r)' + 
\nonumber
\\
& \frac{4 G_0(
M_0 r + 2 M_0 r^2 \epsilon - 4 G_0 J_0^2 \epsilon Y(r)
)
}{r^2(1 + r \epsilon)} 
Y(r)' ,
\end{align}
where
\begin{align}\label{Yder}
Y(r) \equiv 1 - 2 r \epsilon + 2 (r \epsilon)^2 \ln \bigg(1 + \frac{1}{r \epsilon}\bigg).
\end{align}
This solution involves five constants of integration, which are labeled 
$\{ G_0, J_0, M_0, \Lambda_0=-1/\ell_0^2,$ and $\epsilon \}$.
Their naming  and physical meaning  is given from their interpretation 
in two complementary limits.
First, the constant $J_0 \rightarrow 0$ does not appear in the scale dependent
but non-rotating case \cite{Koch:2016uso}. Thus, one imposes that 
for $J_0 \rightarrow 0$ the solution \eqref{setsoluII} reduces to the solution
reported in \cite{Koch:2016uso}, namely
\begin{align}
\lim_{J_0 \rightarrow 0}  G(r) & = \frac{G_0}{1 + r \epsilon},
\\
\lim_{J_0 \rightarrow 0}   N(r) & =  0,
\\
\lim_{J_0 \rightarrow 0}   f(r) & = -\ 8 M_0 G_0 Y(r) + \frac{r^2}{\ell_0^2} ,
\\
\lim_{J_0 \rightarrow 0}   \Lambda(r) & = - \ \frac{r + 3r^2 \epsilon + 8 G_0 \ell_0^2 M_0 \epsilon Y(r) }{\ell_0^2 r (1 + r \epsilon)} 
\nonumber
\\
& \hspace{0.4cm}+ \frac{4 G_0(
M_0 r + 2 M_0 r^2 \epsilon )
}{r^2(1 + r \epsilon)} 
Y(r)' .
\end{align}
The second limit is 
 the rotating classical solution (referring to constant couplings as in \eqref{classical}),
which is obtained when the running paramter $\epsilon$ is taken to be zero,
\begin{align}
\lim_{\epsilon \rightarrow 0} G(r) &= G_0,
\\
\lim_{\epsilon \rightarrow 0} N(r) &= N_0(r) \equiv -\frac{4 G_0 J_0 }{r^2},
\\
\lim_{\epsilon \rightarrow 0} f(r) &= f_0(r) \equiv -8 M_0 G_0 + \frac{r^2}{\ell_0^2} + \frac{16 G_0^2 J_0^2}{r^2},
\\
\lim_{\epsilon \rightarrow 0} \Lambda(r) &= \Lambda_0.
\end{align}
Moreover, when $\{\epsilon, M_0 \} \rightarrow \{0, -1/8G_0 \}$ the appropriate vacuum of the theory is $AdS_{3}$ which is invariant under perturbations due to the running of the couplings controlled by $\epsilon$. 
Further asymptotic corrections can be seen from (\ref{f_asint}).
Since corrections due to quantum scale dependence should be small, 
it is useful to expand the solutions  around $\epsilon\approx 0$
\begin{align}
G(r) &= G_0 \Bigl[1 -r \epsilon + \mathcal{O}(\epsilon^2)\Bigl],
\\
N(r) &= N_0(r) \Bigl[1 - 2 r\epsilon  + \mathcal{O}(\epsilon^2)\Bigl]
\\
f(r) &= f_0(r) + 16 \Bigg[G_0 M_0 - \frac{4 G_0^2 J_0^2}{r^2}\Bigg] r\epsilon + \mathcal{O}(\epsilon^2),
\\
\Lambda(r) &= \Lambda_0 \Bigl[1 + 2 r\epsilon  + \mathcal{O}(\epsilon^2) \Bigl].
\end{align}
 Making this expansion one assumes that the dimensionfull quantity $\epsilon$
is much smaller than any other dimensionfull quantity, such as $r$, $G_0$, $J_0$, or $\Lambda_0$.
In order to get an intuition on the radial dependence of the lapse function $f(r)$ and the corresponding asymptotic behavior one can also refer to a graphical analysis, which is done in figure \ref{figfr} which shows  the lapse function $f(r)$ for different values of $\epsilon$
in comparison to the classical BTZ solution.
One observes that the lapse function $f(r)$ presents two real valued horizons 
after the inclusion of non-zero angular momentum, just like the classical case.
However, the location of those two horizons changes due to the
inclusion of scale dependence.
Thus, for non vanishing $J_0$, there are two horizons independent of the presence ($\epsilon \neq 0$) or absence ($\epsilon=0$) of scale
 dependence.
One remembers that for vanishing angular momentum, there is only a single  horizon for the BTZ black hole which also gets 
shifted to lower values if one allows for scale dependence $\epsilon>0$  \cite{Koch:2016uso}.
 In the scale dependent case there 
 does not exist any finite $\epsilon$ value for which the black hole becomes extremal. 
 This  will be discussed in more detail in section \ref{Thermodynamic}. 
 However, if one considers the limit $\epsilon \rightarrow \infty$, the lapse function approaches that of an extremal black hole.

It is important to note that, some 
relevant quantities, such as the black hole radius $r_H$, depend on the 
scale dependence parameter $\epsilon$. However, 
the asymptotic space-time for $r\rightarrow \infty$ does not show this dependence.
This important fact  will be discussed in more detail in section \ref{Invariants}.

\subsection{Horizon structure}

\begin{figure}[ht!]
\centering
\includegraphics[width=\linewidth]{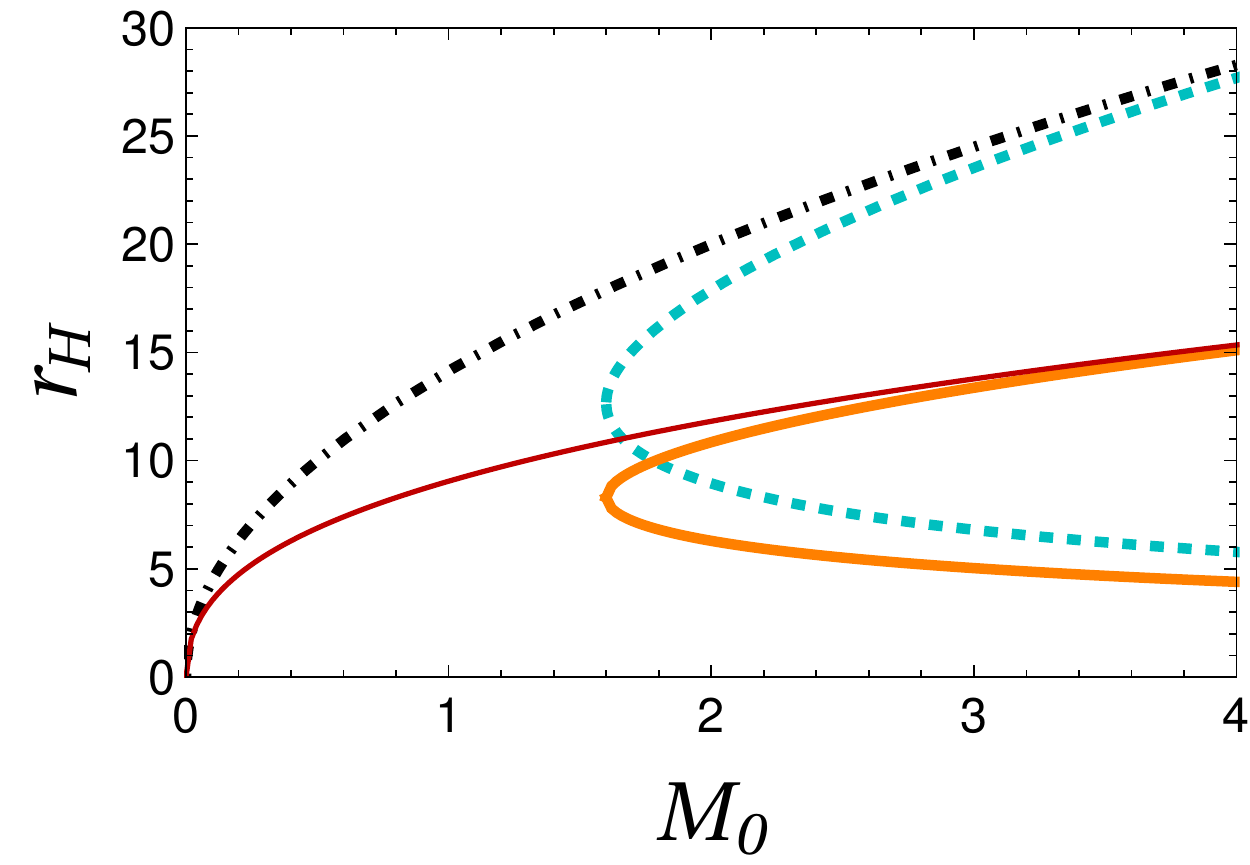}
\caption{\label{rh} 
Black hole horizons $r_H$ as a function of the mass $M_0$ for 
 $\epsilon=0$ and $J_0 =0$  (dotted dashed black line), 
 $\epsilon=0$ and $J_0 = 8$ (blue dashed line), $\epsilon=0.1$ and $J_0 = 0$ (solid thin red line) and $\epsilon=0.1$ and $J_0 =8$ (solid thick orange line). In addition $\ell_0 = 5$ and the values of the rest of the parameters have been taken as unity.
}
\end{figure}

The appearance of  horizons is the defining  criterium 
justifying that solution can be called black hole solution.
The event horizons are defined by $f(r_H)=0$, 
which  can be written as the solutions
of the equation
\begin{align}\label{trans}
Y(r_H) &= \frac{1}{4}\frac{M_0}{G_0 J_0^2} \Bigl[1 \pm \Delta \Bigl] r_H^2
\end{align}
where $\Delta$ remains the same definition given in Eq. \eqref{Delta}
Unfortunately,  this condition has no
closed analytical solution for the scale--dependent lapse function~(\ref{solf}).
Therefore, one has to restrict to a numerical analysis of the black hole horizons
and of the related subjects.
 Fig. \ref{rh} shows the dependence of the horizons $r_H$ on the
 classical mass parameter $M_0$.
One observes that for vanishing angular momentum $J_0=0$ there is only one
real valued horizon with and without scale dependence $\epsilon$.
For finite angular momentum $J_0\neq 0$ there appears a second inner horizon.
In all studied cases, the effect of the scale dependence $\epsilon>0$ was to reduce the outer horizon radius with respect to the non-scale dependent case $\epsilon=0$.
Even though the analytical solution for the horizon is not obtained, 
one still can analyze the lapse function in a regime when the $\epsilon$ correction is small. 
The event horizon, up to leading order, is
\begin{align}
r_H &\approx r_0 \Bigl[1 - \epsilon r_0 + {\mathcal{O}}(\epsilon^2)\Bigl],
\end{align}
where one indeed observes the expected deviation of the horizon with respect the classical case. 
One notes that in the scale--dependent scenario the event horizon decreases when $\epsilon > 0$ or increases when $\epsilon < 0$. 
This feature reveals that the black hole thermodynamics is directly affected. 

For the inner horizon and for large values of $M_0$, the lapse function takes an simplified form, which allows to express the horizon as
\begin{align}
r_H^0 &= \sqrt{\frac{2 G_0}{M_0}}J_0 \bigg[ 1 - 2 \bigg( 
\sqrt{\frac{2 G_0}{M_0}}J_0
\bigg) \epsilon  + \mathcal{O}(\epsilon^2)\bigg],
\end{align}
where one recovers the classical horizon in the limit $\epsilon \rightarrow 0$.

\section{Invariants and asymptotic space-times}\label{Invariants}
  
This section discusses different asymptotic limits.
In particular, we will focus on the asymptotic line element and the behavior of the the Ricci scalar $R$.

\subsection{Asymptotic line element}

\subsubsection{Behaviuor when $r \rightarrow 0$}

When we are close to the horizon, the lapse and shift functions suffer deviations respect the classical solution. 
In order to emphasize that, we expand our result around $r$ up to first order to get
\begin{align}
\mathrm{d}s_{0^+}^2 &= -f_{0^+} \mathrm{d}t^2 + f_{0^+}^{-1} \mathrm{d}r^2 + r^2 \left[ N_{0^+} \mathrm{d}t + \mathrm{d}\phi \right]^2 ,
\end{align}
with
\begin{align}
\begin{split}
f_{0^+}(r) = & - 8 M_0 G_0 \bigl[1- 2 \epsilon r \bigl] 
\\
& + \frac{16 G_0^2 J_0^2}{r^2} \bigg[1 - 4 \epsilon r\bigg] + \mathcal{O}(r^2), \label{fshort}
\end{split}
\\
N_{0^+}(r) &= N_0(r) \bigl[1 - 2 \epsilon r  + \mathcal{O}(r^2)\bigl],
\end{align}
where we only are considering terms up to linear order in $\epsilon$. Given these expressions, it is very obvious that the lapse and shift functions decreases if $\epsilon > 0$, respect the usual solution.

\subsubsection{Behaviuor when $r \rightarrow \infty$}
 The asymptotic line element is expressed in terms of asymptotic lapse and shift function (at large radii respect to the inverse of scale dependent parameter), i.e.
\begin{align}
\mathrm{d}s_{\infty}^2 &= -f_{\infty} \mathrm{d}t^2 + f_{\infty}^{-1} \mathrm{d}r^2 + r^2 \left[ N_{\infty} \mathrm{d}t + \mathrm{d}\phi \right]^2 .
\end{align}
where the aforementioned functions are shown below
\begin{align}
f_{\infty}(r) &= \frac{r^2}{\ell_0^2} - 8 M_0 G_0 \left(\frac{2}{3}\frac{1}{r \epsilon}\right) + \mathcal{O}\left(\frac{1}{r^2}\right), \label{f_asint}
\\
N_{\infty}(r) &= N_0(r)\left(\frac{2}{3}\frac{1}{r \epsilon}\right) + \mathcal{O}\left(\frac{1}{r^4}\right). \label{N_asint}
\end{align}
It is important to note that the asymptotic lapse function mimics at leading order an $AdS_3$ behavior. Going further,
we observe that the lapse function given in Eq.~\eqref{f_asint}, at sub--leading order, 
reflects the effect of the scale--dependent scenario through a factor $1/(\epsilon r)$.
For the shift function, the scale dependent effect is dominant at leading order in $r$ (which is given in Eq. \eqref{N_asint}), 
which means that asymptotically the running of the gravitational coupling modifies the classical behavior. In addition, the quantum correction in both functions appear as a term $\sim 1/ (r\epsilon)$.
Regarding the lapse function, if one remains only the dominant term in the large radius limit, the asymptotic structure does not change, therefore, it is equivalent to $AdS_3$, which is consistent with our previous work \cite{Koch:2016uso}. 

When studying the sub-leading corrections
one has to be carefull with the two competing limits
$\epsilon \rightarrow 0$ and $r \rightarrow \infty$, which can not be commuted.
In this context we note that the naming of the integration constants ($J_0, M_0, \dots$) 
and thus of their physical interpretation was based on the classical limit $\epsilon \rightarrow 0$.

\subsection{Asymptotic Invariants}

For the study of coordinate independent properties of a solution it is useful to refer to invariants.
For the given metric \eqref{lineelans} the Ricci scalar is given below
\begin{align}
R &= \frac{1}{2 r}\left(r^3 N'(r)^2-4 f'(r)\right) - f''(r),
\end{align}
which, after explicit insertion, reads as follows
\begin{align}
\label{R_full}
R = & R_0 + 16 G_0 M_0 \frac{Y'(r)}{r} 
\bigg[ 
1
+ 
\frac{2 G_0 J_0^2}{M_0} \frac{Y(r)}{r^2} -
\\
& \frac{3 G_0 J_0^2}{2M_0} \frac{Y'(r)}{r} \bigg]
+ 8 M_0 G_0 Y''(r)\left[1-\frac{4 G_0 J_0^2}{M_0} \frac{Y(r)}{r^2}\right]
\nonumber
\end{align}
From Eq. \eqref{R_full} we get the classical solution after demand that $\epsilon \rightarrow 0$, which reads
\begin{align}
R_0 &\equiv 6 \Lambda_0,\label{classical_R}
\end{align}

\subsubsection{Behaviuor when $r \rightarrow 0$}

For small $r$ the invariant  expansion of Eq. \eqref{R_full} 
gives
\begin{align}\label{seriesR}
R &=  -\frac{64 G_0^2 J_0^2}{r^3} \epsilon 
\left( 1+ \mathcal{O} (r)\right).
\end{align}
One observes that the presence of scale--dependent couplings ($\epsilon \neq 0$) produces a singularity at $r = 0$.
This finding is somewhat surprising since one might have hoped that 
quantum induced scale dependence would help with singularity problems of the classical theory and not make them worse.
However, the implementation of scale dependence that was used here is clear and determinating the solution under the given assumptions.
Thus, one has to conclude that the solution of the singularity problem shown in (\ref{seriesR}) has to come
from a framework that falls outside of our assumptions such as a line element with different structure, or the addition of non-local or
higher order terms in the effective action.

\subsubsection{Behaviuor when $r \rightarrow \infty$}

The other asymptotic regime of interest is the
large radius expansion $r\to\infty$. 
In this regime one can approximate the logarithm contribution according to $\ln(1+z) \approx z - z^2/2$ (using $z = 1/\epsilon r$). In this limit the Ricci scalar is given by
\begin{align} \label{Ricci_Lejos}
R &= R_0 - \frac{32 M_0 G_0 }{\epsilon r^3} + \mathcal{O}\bigg(\frac{1}{r^4}\bigg).
\end{align}
Please note that  the Ricci  is asymptotically finite
independent of the order one takes the competing limits $r \rightarrow \infty$ and $\epsilon \rightarrow 0$.
However, due to the expansion of the logarithms, the expression \eqref{Ricci_Lejos} is only valid if $r\gg 1/\epsilon$.
As we know, the Ricci scalar is constant in the classical case \eqref{classical_R} and therefore for certain values of the parameter $\epsilon$, asymptotically the Ricci scalar is well-behaved take the classical value~(\ref{classical_R}). 

\section{Thermodynamic properties} \label{Thermodynamic}
The (numerical) knowledge of the horizons allows to study the
thermodynamic properties of the scale dependent rotating black hole solution \eqref{solf}.

\subsection{Hawking Temperature}
\begin{figure}[ht!]
\centering
\includegraphics[width=\linewidth]{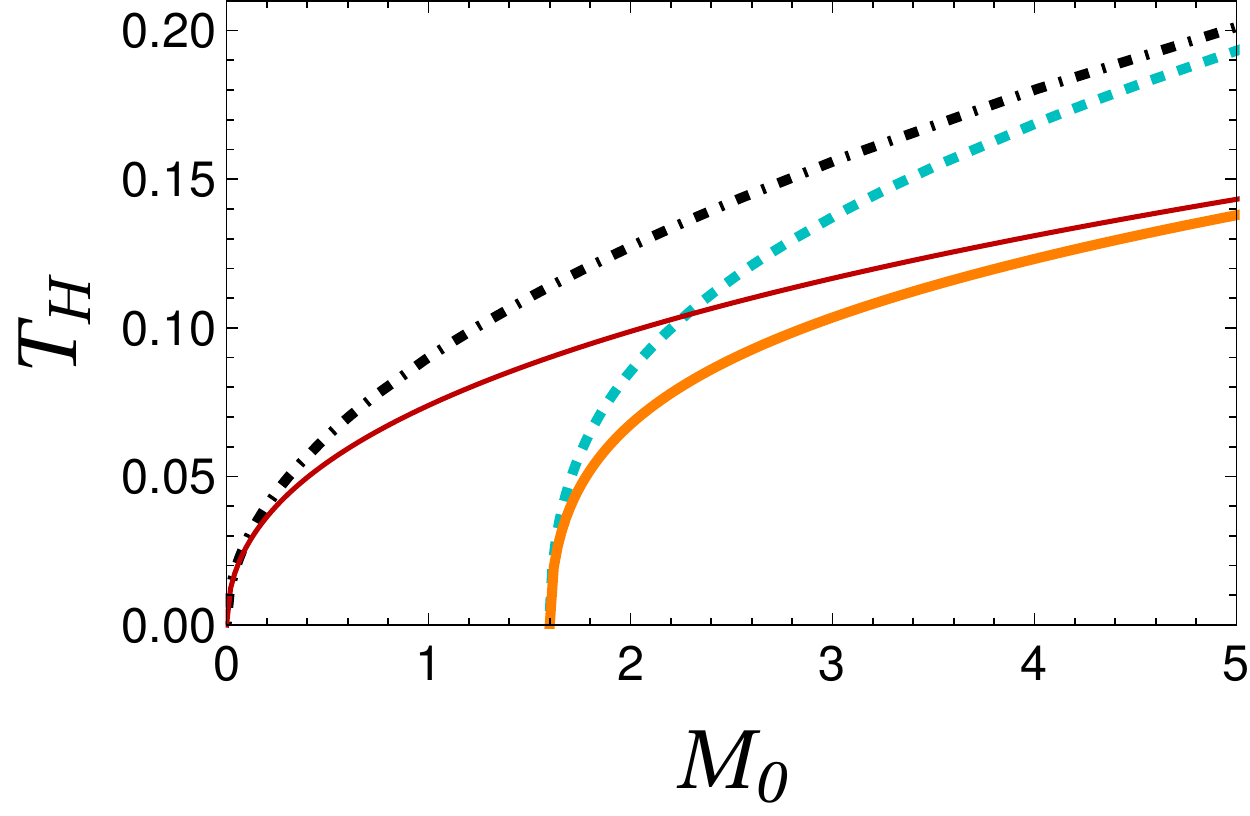}
\caption{\label{TH_0} 
The Hawking temperature $T_H$ as function of the classical mass $M_0$ for four different cases:  $\epsilon=0$ and $J_0 =0$  (dotted dashed black line), $\epsilon=0$ and $J_0 = 8$ (blue dashed line), $\epsilon=0.1$ and $J_0 = 0$ (solid thin red line) and $\epsilon=0.1$ and $J_0 =8$ (solid thick orange line). In addition $\ell_0 = 5$ and the values of the rest of the parameters have been taken as unity.
}
\end{figure}

The Hawking temperature of a black hole assuming a circularly symmetric line element (\ref{lineelans}), is defined by
\begin{align}
T_H(r_H) &= \frac{1}{2 \pi} \left| \frac{1}{2}\frac{\partial f}{\partial r} \bigg|_{r=r_H}\right|,
\end{align}
which gives for the solution of (\ref{solf})
\begin{align}\label{THH}
T_H(r_H) &= \frac{1}{4 \pi}\Bigg|\frac{16 M_0 G_0 }{r_H(1 + \epsilon r_H) } \Delta \Bigg|.
\end{align}
Please, note that 
this formula coincides with the classical expression, if one
replaces $G_0$ by $G(r_H)$ in Eq. \eqref{Tclasica}.
As it can be seen from (\ref{THH}), the Hawking temperature vanishes for $\Delta=0$.
The extremal black hole is given when $M_0 \ell_0$, which is the same extremality condition as 
in the classical case (\ref{Jcritico}).
Figure~\ref{TH_0} shows the  temperature which takes into account the running coupling effect
in comparison to the ``classical'' temperature,
as a function of the parameter $M_0$.
We notes that indeed the curves with ($\epsilon\neq 0$) and without scale dependence ($\epsilon=0$)
coincide at the same minimal mass $M_0=J_0/\ell_0$.

Since scale dependence is motivated by quantum corrections
and since those corrections are typically small,
it can be expected that
 the integration constant $\epsilon$, which parametrizes the scale dependence,  
 is small.
Under this assumption one can expand for $r \epsilon \ll 1$ to get the well-known
Hawking temperature (at leading order) i.e.
\begin{align}
T_H(r_H^0) &= T_0(r_{0^+}) \bigl| 1 + 4 r_{0^+} \epsilon + \mathcal{O}(\epsilon^2)\bigl|
\end{align}
where $r_{0^+}$ is the classical horizon $r_H$ which is a solution of (\ref{classical}) evaluated when $r$ is close to zero. We wish to remark that this approximation is used because we always assume a weak coupling $\epsilon$. Besides, the classical Hawking temperature $T_0(r_{0^+})$ is computed following the usual procedure for the lapse function (\ref{classical}) when $r$ is small.

\subsection{ Bekenstein-Hawking entropy }
The Bekenstein-Hawking entropy is also valid for theories in which the gravitational coupling is 
variable  \cite{Jacobson:1993vj,Iyer:1995kg,Visser:1993nu,Creighton:1995au}. 
For black hole solutions in $D +1$ dimensions with varying Newton's
coupling the entropy is given by
\begin{align}
S &= \oint {\mathrm {d}}^{D-1} r    \frac{\sqrt{h}}{4 G(r)},
\end{align}
where $h_{ij}$ is the induced metric at the horizon $r = r_H$.
 For the present circularly symmetric solution the aforementioned integral is straightforward. The induced line element 
 for constant $t$ and $r$ slices is simply ${\mathrm {d}}s = r{\mathrm {d}}\theta$ and moreover $G_H = G(r_H)$ is constant along the horizon. Therefore, the entropy for the solution (\ref{solf}) is
\begin{align}\label{entropy_full}
S &=\frac{\mathcal{A}_H(r_H)}{4G(r_H)} = S_0(r_H)(1 + \epsilon r_H).
\end{align}

\begin{figure}[ht!]
\centering
\includegraphics[width=\linewidth]{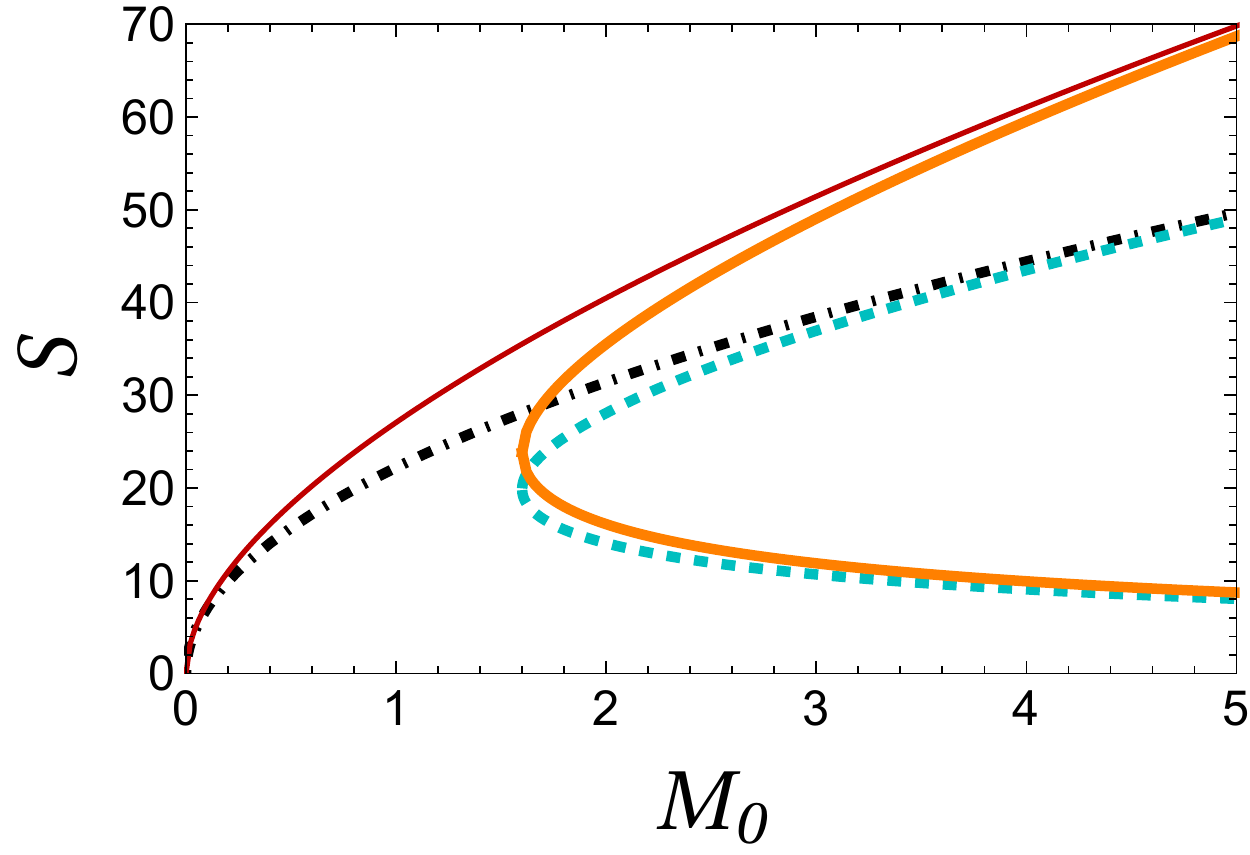}
\caption{\label{S_0} 
The Bekenstein-Hawking entropy $S$ as function of classical mass $M_0$ for four different cases: 
 $\epsilon=0$ and $J_0 =0$  (dotted dashed black line), $\epsilon=0$ and $J_0 = 8$ (blue dashed line), $\epsilon=0.1$ and $J_0 = 0$ (solid thin red line) and $\epsilon=0.1$ and $J_0 =8$ (solid thick orange line). In addition $\ell_0 = 5$ and the values of the rest of the parameters have been taken as unity.
}
\end{figure}

Figure \ref{S_0} shows 
the entropy for our BTZ rotating scale--dependent black hole as a function of $M_0$. 
We observe that when $J_0=0$ both, 
the classical entropy ($\epsilon=0$) and the scale--dependent entropy ($\epsilon\neq 0$) tend to zero for $M_0 \rightarrow 0$, whereas for $J_0 \neq 0$ both, the classical and the scale--dependent solution, present a cut-off for the critical mass $M_0 =J_0/\ell_0 $.
An analytic expression can be can be obtained in certain limit.
By considering small values of $\epsilon$ it is possible to expand this
expression
\begin{align}
S(r_H^0) &= S_0(r_H^0) \left[1 + \epsilon r_H^0 +\mathcal{O}(\epsilon^3) \right].
\end{align}
Thus, the quantum effect increases the entropy respect the classical solution. 

\section{Discussion}\label{Discussion}
Effective quantum corrections can be systematically introduced to the BTZ black hole by
assuming a scale--dependent framework.
This implies non-trivial deviations from classical black hole solutions. 
In this work, one of the integration constants ($\epsilon$) of the generalized field equations   is used
as a control parameter, which allows to regulate the
strength of scale dependence, such that for $\epsilon \rightarrow 0$, the 
well-know classical BTZ background is recovered.
This article discusses the BTZ black hole taking into account angular momentum in the context of scale dependent couplings.
A solution of the corresponding field equations is presented
and compared it with three different known cases: the classical case ($\epsilon=0$) without angular momentum,
 the classical case ($\epsilon=0$) with angular momentum,
 and the scale dependent case ($\epsilon \neq 0$) without angular momentum. 

The new scale--dependent solution has some interesting features, for instance the lapse function increases rapidly when $r \rightarrow \infty$ (which is present in the classical case) but now the effect is deeper, see Fig. \ref{figfr} and compare the black curve ($\epsilon = 0$) with red curve ($\epsilon = 1$). 
By comparing Eq. \eqref{classical} with Eq. \eqref{f_asint} and with Eq. \ref{fshort}, we observe the deviation given by the scale--dependent framework respect to the classical solution. It is remarkable that when we are close to the origin the lapse function suffers a shift, while when we are far from the origin it shows a decrease by a factor of $1 / \epsilon r$. In both cases the solution is affected.

Furthermore, according to Fig, \ref{rh}, the outer horizons decrease when $\epsilon$ increases.
The effect of the scale dependent approach is thus that it produces smaller horizons, when compared to the usual case. Interestingly this decrease does not come with a change of the critical mass, where the two outer horizons merge.

An analysis of the Ricci scalar reveals that a singularity appears at $r\rightarrow 0$ which is absent in the corresponding classical BTZ solution.
Indeed, the BTZ black hole has a constant scalar, according to Eq. (\ref{classical_R}), whereas in the scale dependent case 
($\epsilon \neq 0$) the singularity at $r=0$ is always present according with Eqs. (\ref{seriesR}). This is a consequence of the scale--dependent scenario.

Regarding the Hawking temperature, it is interesting that the scale dependent formula and the corresponding classical counterpart, coincide, under the replacement $G_0 \rightarrow G(r_H) = G_0/(1+ \epsilon r_H)$ \eqref{setsoluII}.
It is further remarkable that the extreme black hole condition is also maintained and, therefore, the Hawking temperature is equal to zero when $M_0 ^{\text{min}}=J_0/\ell_0$, independent of the strength of scale dependence $\epsilon$. 
Moreover, we note that in presence of scale--dependent couplings the temperature is lowered with respect to the classic BTZ solution for large values of $M_0$. 
Whereas when $M_0$ is close to zero (for $J_0 = 0$) and when $M_0$ is close to $M_0 ^{\text{min}}$ (for $J_0 \neq 0$), the classical and the scale dependent solution 
are very similar.
One notes that the Bekenstein-Hawking entropy is increased by the scale dependence $\epsilon \neq 0$ and that
for large values of $M_0$ the solutions with and without angular momentum match for a given value of $\epsilon$,
but they differ for different values of $\epsilon$.
Throughout the numeric analysis we also have used a relatively
 ``small" value of $\epsilon$, a choice which can be motivated by the assumption of relatively weak quantum effects provoking
 scale dependence at the level of the effective action (\ref{actionEG}).
Lets mention in this context that the integration constant $\epsilon$ can be made dimensionless for example
by defining $\epsilon= \bar \epsilon M_0$, in which case the graphical and analytical results with respect to $\bar \epsilon$
would have to be rescaled correspondingly.

Finally, lets comment on the ansatz (\ref{lineelans}).
This type of ansatz also works for the spherically symmetric case.
However, inspired by the ideas presented by Jacobson \cite{Jacobson:2007tj}
it was possible to show that, for spherically symmetric static black holes,
this type of ansatz is actually a consequence of
a simple Null Energy Condition (NEC) \cite{Koch:2016uso,Rincon:2017goj,Rincon:2017ypd}.

This condition allows the avoidance of pathologies such as tachyons, instabilities, and ghosts \cite{Wald:1984rg,Rubakov:2014jja}.
Further, the NEC plays a crucial role in the Penrose singularity theorem \cite{Penrose:1964wq}.
However, a straight forward implementation of a generalized NEC to the rotating BH was not achieved, since the appearance of angular momentum reduces
the symmetry of the problem. One would first have to generalize the arguments 
given in \cite{Jacobson:2007tj} to the rotational symmetry, before one
can try to build an argument deriving the ansatz (\ref{lineelans}), 
as a consequence of some kind of NEC.
Thus, at this point the use of the ansatz (\ref{lineelans}) is 
well justified, since it agrees with the NEC for vanishing rotation and since it further implements the structure of the line element for the case of the classical (not scale-dependent) counterpart.


\section{Conclusion} \label{Conclusion}

In  this  work  we  have  studied  the  scale  dependence
of the rotating BTZ black hole assuming a finite cosmological term in the action. 
After presenting the models and the
classical black hole solutions, we have allowed for a scale
dependence of the cosmological ``constant" as well as the gravitational coupling, and we have solved the corresponding  generalized   field  equations with static circular symmetry.
We have compared the classical solutions distinguishing two different cases, 
i.e. with and without angular momentum,
 with the corresponding scale dependent solution for same values of angular momentum.
In addition, the horizon structure, the asymptotic spacetime and the thermodynamics were analyzed.
 In particular, the analysis of the Hawking temperature allowed to find a extremal black hole  which coincides with the classical counterpart.

\begin{acknowledgements}
We wish to thank Prof. Maximo Ba\~na\-dos for some illuminating comments.
The author A.R. was supported by the CONICYT-PCHA/\- Doctorado Nacional/2015-21151658.
The author B.K. was supported by the Fondecyt 1161150 and Fondecyt 1181694.
\end{acknowledgements}


\bibliographystyle{unsrt}         

\end{document}